\documentclass[a4paper]{jpconf}
\usepackage{graphicx}
\usepackage{url}
\usepackage{needspace}
\usepackage[pdfencoding=auto]{hyperref}
\hypersetup{%
 bookmarksnumbered=true,%
 colorlinks=true,%
 pdftitle={Achievement of 200,000 hours of operation at KEK 7-GeV electron 4-GeV positron injector linac},%
 pdfauthor={Kazuro Furukawa},%
 pdfsubject={IPAC 2022 THPOST012},%
 pdfkeywords={SuperKEKB Injector Linac\000\012200,000 hours of operation\000\012}}
\begin{document}
\title{Achievement of 200,000 hours of operation at\\
KEK 7-GeV electron 4-GeV positron injector linac}

\author{K~Furukawa$^{1,2}$,
M~Akemoto$^1$,
D~Arakawa$^1$,
Y~Arakida$^1$,
Y~Bando$^2$,
H~Ego$^{1,2}$,
Y~Enomoto$^{1,2}$,
T~Higo$^1$,
H~Honma$^1$,
N~Iida$^{1,2}$,
K~Kakihara$^1$,
T~Kamitani$^{1,2}$,
H~Katagiri$^1$,
M~Kawamura$^1$,
S~Matsumoto$^{1,2}$,
T~Matsumoto$^{1,2}$,
H~Matsushita$^1$,
K~Mikawa$^1$,
T~Miura$^{1,2}$,
F~Miyahara$^{1,2}$,
H~Nakajima$^1$,
T~Natsui$^{1,2}$,
Y~Ogawa$^1$,
S~Ohsawa$^1$,
Y~Okayasu$^{1,2}$,
T~Oogoe$^1$,
M~A~Rehman$^1$,
I~Satake$^1$,
M~Satoh$^{1,2}$,
Y~Seimiya$^{1,2}$,
T~Shidara$^1$,
A~Shirakawa$^1$,
H~Someya$^1$,
T~Suwada$^{1,2}$,
M~Tanaka$^1$,
D~Wang$^1$,
Y~Yano$^1$,
K~Yokoyama$^{1,2}$,
M~Yoshida$^{1,2}$,
T~Yoshimoto$^{1,2}$,
R~Zhang$^{1,2}$ and
X~Zhou$^{1,2}$}

\address{$^1$ High Energy Accelerator Research Organization (KEK), Tsukuba, Ibaraki, 305-0801, Japan}
\address{$^2$ Graduate University for Advanced Studies (SOKENDAI), Tsukuba, Ibaraki, 305-0801, Japan}

\ead{kazuro.furukawa@kek.jp}

\begin{abstract}
KEK electron positron injector LINAC initiated the injection operation into Photon Factory (PF) light source in 1982.  Since then for 39 years, it has served for multiple projects, namely, TRISTAN, PF-AR, KEKB, and SuperKEKB.  Its total operation time has accumulated 200 thousand hours on May 7, 2020.  We are extremely proud of the achievement following continuous efforts by our seniors.  The construction of the injector LINAC started in 1978, and it was commissioned for PF with 2.5 GeV electron in 1982.  In parallel, the positron generator linac was constructed for the TRISTAN collider project.  The slow positron facility was also commissioned in 1992.  After the KEKB asymmetric-energy collider project was commissioned in 1998 with direct energy injections, the techniques such as two-bunch acceleration and simultaneous injection were developed.  As the soft structure design of the LINAC was too weak against the great east Japan earthquake, it took three years to recover.  Then the construction and commissioning for the SuperKEKB project went on, and the simultaneous top-up injection into four storage rings contributes to the both elementary particle physics and photon science.
\end{abstract}

\section{Introduction}

In the experimental particle physics research in the 1970s, there was growing expectation for the world-class Japanese domestic collider after the successes in the world-level theoretical research. On the other hand, in the field of synchrotron radiation science, there were demands for the establishment of a synchrotron radiation research institute with a dedicated accelerator. 
In order to realize both of those accelerator projects, a 400-m electron linear accelerator capable of 2.5 GeV direct energy injection was constructed in 1982. 

\begin{figure*}[t]
\centering
\includegraphics[width=0.95\textwidth]{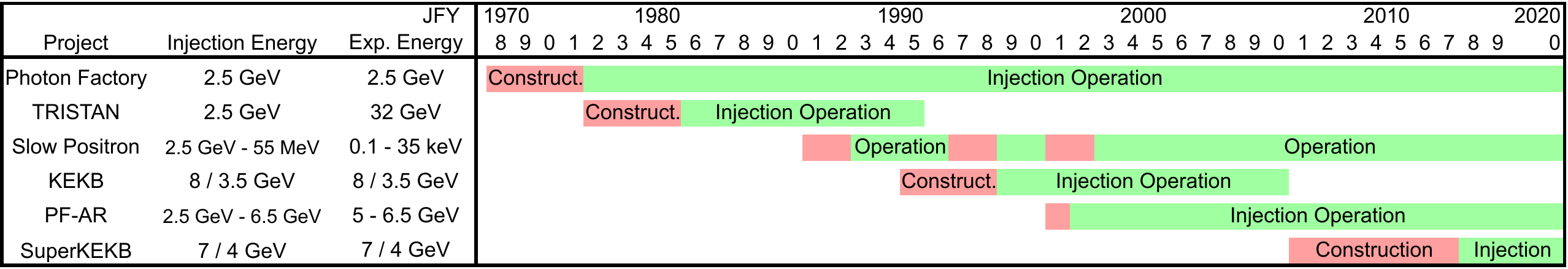}
\caption{
Electron positron accelerator projects operated by the injector LINAC with their injection beam energy and experimental beam energy.
}
\label{projects}
\end{figure*}

Since then, the injector had been in operation for 38 years in 2020, supporting the consecutive accelerator projects of Photon Factory (PF), TRISTAN, KEKB, PF-AR, and SuperKEKB as depicted in 
Figs. \ref{projects}, \ref{tristan-config}, \ref{kekb-config} and \ref{skekb-config}. It has finally achieved 200,000 hours of operation 
in May 2020. The brief operation history would be given.

\section{2.5 GeV injection for PF/TRISTAN}


At the time of the initial construction since 1978, several university-based accelerator facilities had achieved research successes in Japan. It was fortunate that a new linear electron accelerator had been built to share resources for advanced scientific research in both particle physics and synchrotron radiation science, which were based on the accumulated experiences at universities. 
The situation might be similar to recent J-PARC's focus on both nuclear and particle physics, as well as neutron and muon science. It seems that there are not many accelerator projects in the world that have multiple disciplines from the beginning. Those Japanese accelerators are quite unique that deliver beams for multiple science fields.  

\begin{figure}[b]
\centering
\includegraphics[width=0.72\columnwidth]{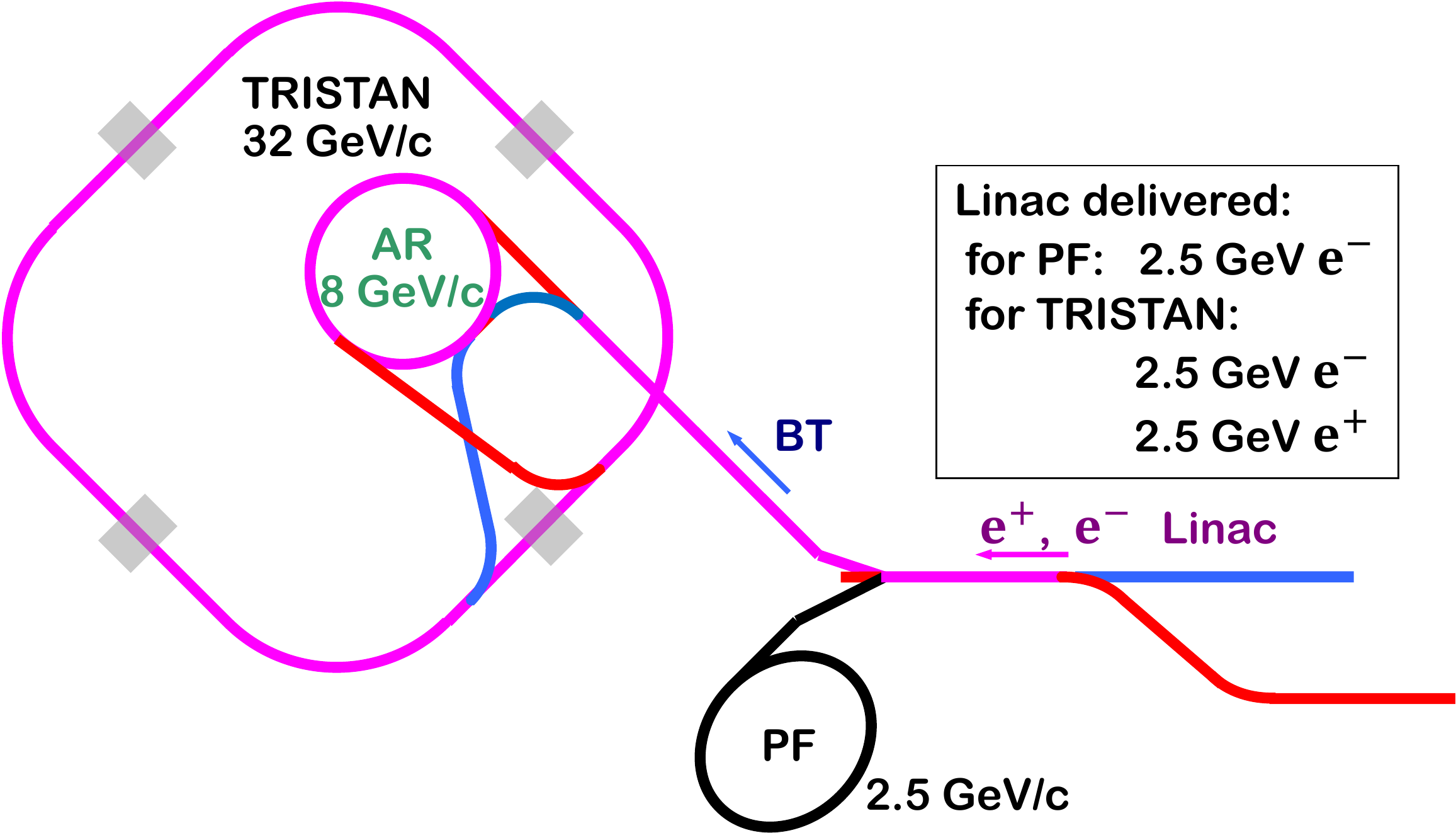}
\caption{
Injection for TRISTAN and PF.
}
\label{tristan-config}
\end{figure}

While many researchers of synchrotron radiation research wanted the institute independent from particle physics, the accelerator science played a role as a bridge between those two.
One hundred sixty of 
2-m-long S-band traveling-wave accelerating structures with $2\pi/3$ mode quasi-constant gradient were installed, and driven by forty 20-MW klystrons to achieve 2.5 GeV electron and positron beams~\cite{linac-1980}. 2.5 GeV direct energy injection was made for PF and 2.5 GeV electron positron beams from the injector were accelerated through the TRISTAN synchrotron chain for the 32 GeV collision as depicted in Fig.~\ref{tristan-config}.

When the synchrotron radiation facility, PF, was completed, there must be an enthusiastic response to the dedicated synchrotron radiation facility that would promote research in synchrotron radiation science. The TRISTAN project also began to take the lead in high-energy experimental physics, after Japanese theorist had contributed to the worldwide development of particle physics theory.  It was realized recently that the specifications of the electric power facilities built at that time secured the expansion possibility even for the fields of nuclear physics and ultraviolet synchrotron radiation science.  Such design facilitated the upgrades towards KEKB and SuperKEKB later.

\section{8-GeV upgrade for KEKB}


Since the construction of TRISTAN, a linear collider was expected as the next particle physics experiment project.  Instead of immediately promoting a large scale collider, however, the construction of a B Factory, KEKB, was carried out to experimentally confirm the remaining issues of CP violation in particle theory proposed by Makoto Kobayashi and Toshihide Maskawa. Since the injection performance was important to achieve the integral performance as a factory, there were concerns compared to SLAC/PEP-II, which was planning the B-factory in parallel. 

\begin{figure}[b]
\centering
\includegraphics[width=0.72\columnwidth]{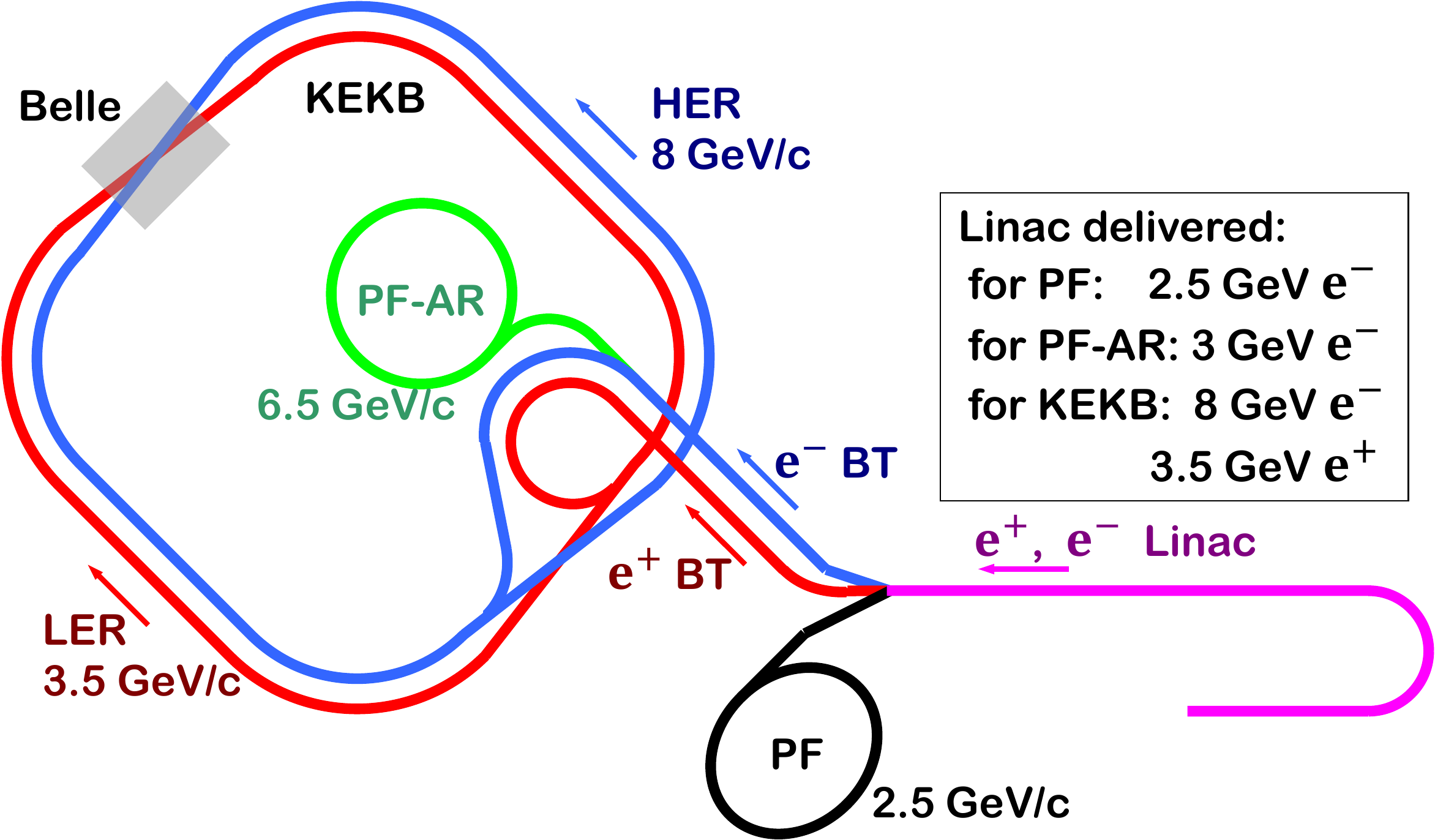}
\caption{
Injection for KEKB, PF and PF-AR.
}
\label{kekb-config}
\end{figure}

The direct energy injections for the asymmetric energy collider rings were realized by an energy upgrade from 2.5~GeV to 8~GeV by introducing SLED microwave pulse compressors and adding 50\% more accelerating structures that extends the length of the injector up to 700 m in a shape of a letter `J' as in Fig.~\ref{kekb-config}. It was a substantial and difficult construction because the injector continued the PF injection~\cite{linac-kekb1,linac-kekb2}.  It made maximum use of the opportunities of the longest shutdown of nine months in 1997 during the construction.  



The success of the KEKB operation was largely due to the online operating environment of the SAD beam simulation software and the EPICS control framework which were developed during the injector commissioning period~\cite{sad-icap1998}.

KEKB and PEP-II projects developed the friendly competition, and they compared the performance of the accelerators in those control rooms everyday~\cite{pepii-epac2008}.  The performance of the KEK injector was gradually improved during the project step-by-step~\cite{kekb-eefact2018}. 


\begin{figure}[t]
\centering
\includegraphics[width=0.8\columnwidth]{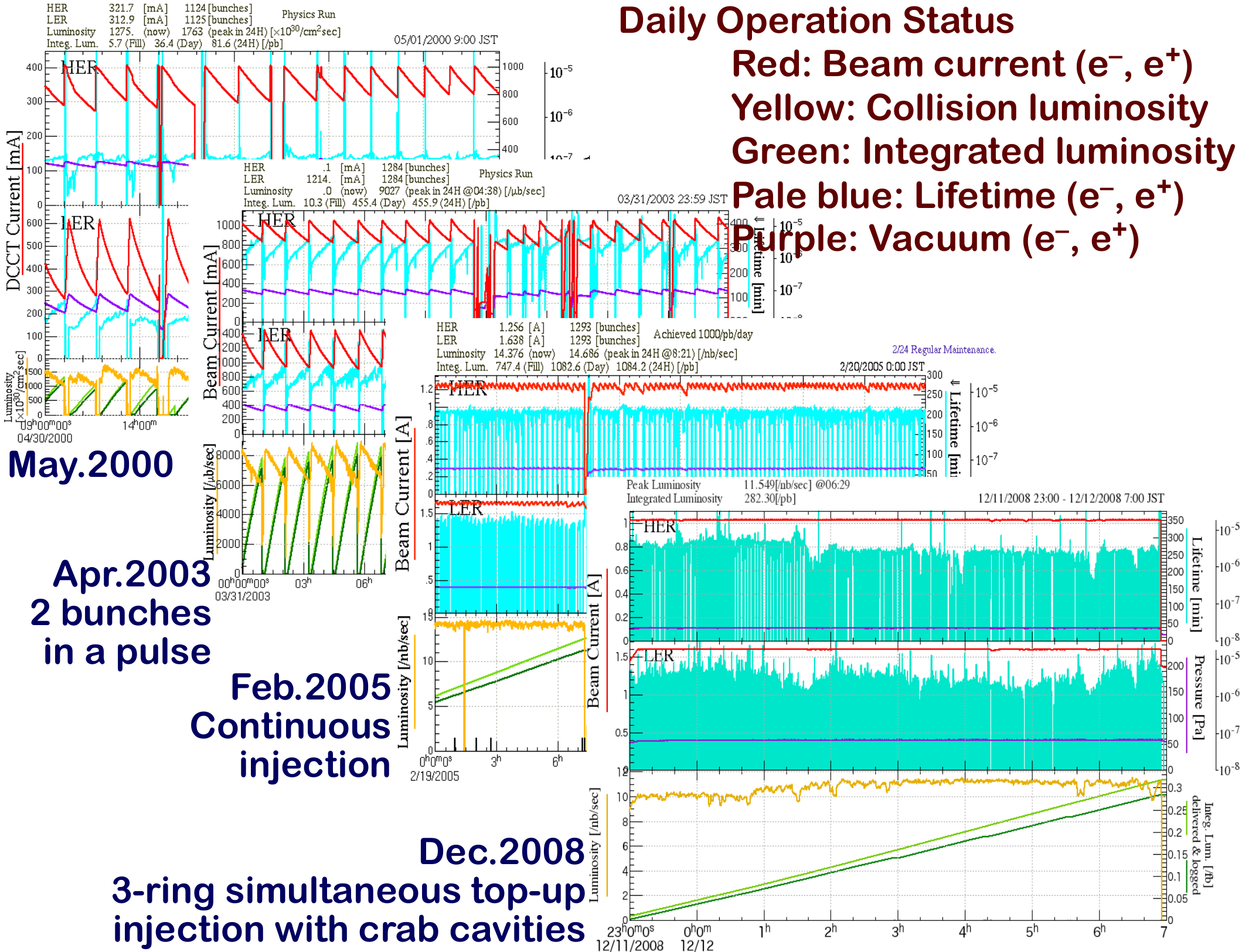}
\caption{
Progress of the daily operation during the KEKB project, with the stored beam current in red and collision luminosity in yellow.
}
\label{kekb-progress}
\end{figure}

The positron beam injection was enhanced by employing dual bunch acceleration.  The separation of two bunches was chosen to be 96 ns because of the frequency relationship between the injector and the ring~\cite{dualbunch-ical2001}. 

Then later, it also succeeded in continuous injection by switching the injection every several minutes while the Belle collision detector was in operation. In 2008, the injection was further improved by the simultaneous top-up injection in to three storage rings except PF-AR in pulse-to-pulse modulation switching beams with distinct properties every 20 ms employing event-based control architecture~\cite{event-ical09} as in Fig.~\ref{kekb-progress}. 

\section{Slow positron and others}

The injector facility usually contributes by injecting beams into the downstream storage rings. However, there are cases where experiments are performed using direct electrons. In the past at KEK LINAC, there have been experiments at the beam dump for a search for an axion like particle, beam irradiation calibration of tile fiber calorimeter detectors for the SSC project, positron beam channeling experiments, and positron production using electron channeling radiation. As a result of the positron generation experiment, a crystal target was actually applied to the positron generation for KEKB injection in 2006, and the positron generation efficiency was successfully improved by about 25\%~\cite{channeling-prab2007}. 

Another ongoing project is the research experiments for material science and particle physics by slow positrons. Between injections to the TRISTAN project slow positrons were generated at the end of LINAC. While the experimental setup had to be moved twice due to the change in the beam operation of the main LINAC, now a dedicated 50 MeV linac is used for slow positron experiments sharing the same resources as the main LINAC~\cite{slowpos-wada2018}. It is valuable to produce scientific results within the LINAC building.

\section{SuperKEKB and simultaneous top-up injection}

After the KEKB project was completed in 2010 the injector, whose girders were built with the concept of a flexible structure, suffered a serious damage due to the Great East Japan Earthquake in 2011. While light source injections were recovered in 3 months, it took three years to start the beam test for SuperKEKB as the alignment was severely deteriorated.  Actually, the alignment requirement became much tight in order to suppress the wakefield effect in the accelerating structure.  
While the light source injections were maintained even during the upgrade towards SuperKEKB, the longest construction period of five months was secured in 2017.  

\begin{figure}[t]
\centering
\includegraphics[width=0.72\columnwidth]{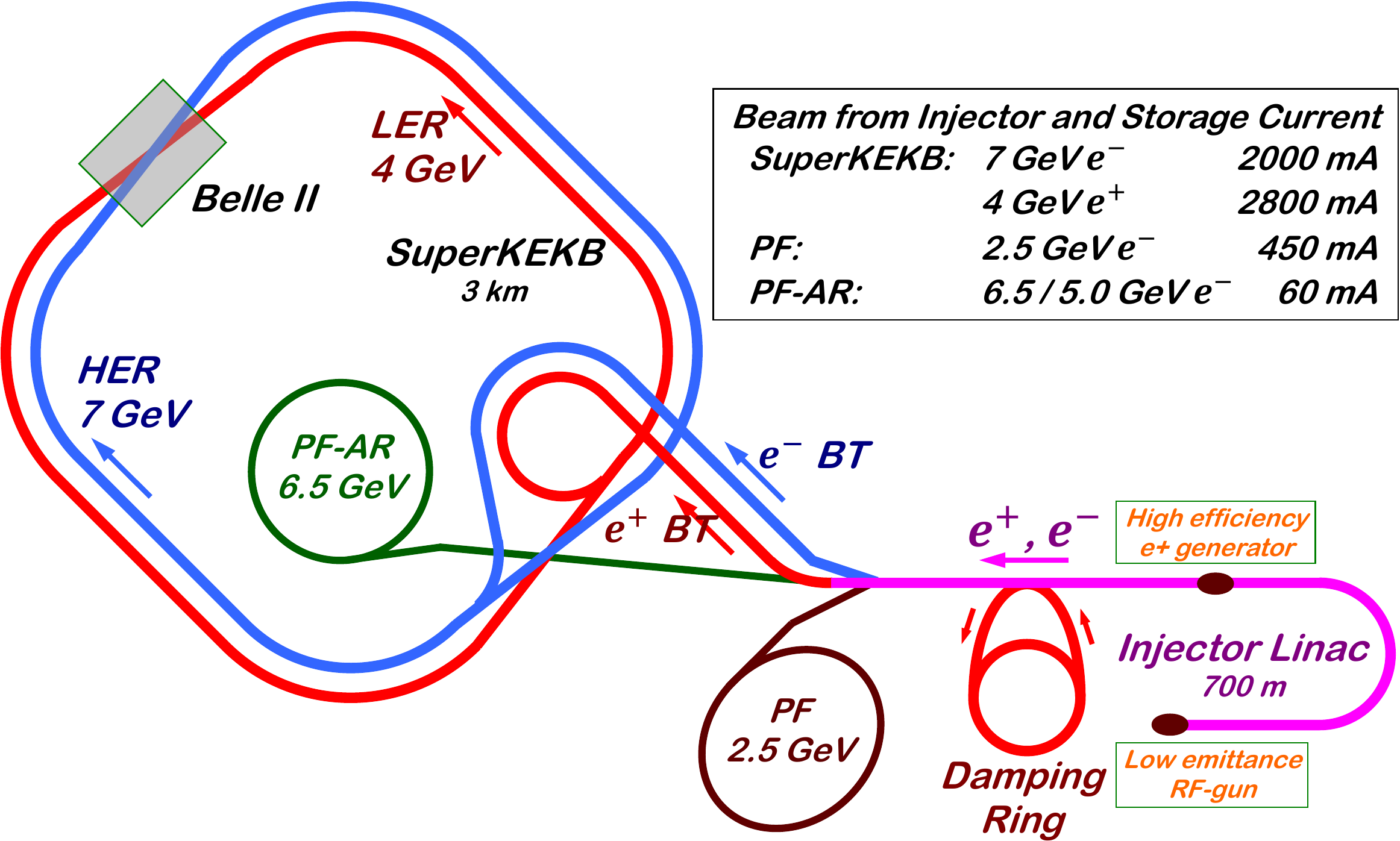}
\caption{
Injection for SuperKEKB as well as PF and PF-AR.
}
\label{skekb-config}
\end{figure}

The energy enhancement was the main subject for the previous KEKB injection upgrade, on the other hand the SuperKEKB project required quality improvements of the injection beam for the electron positron collision with the novel nano-beam scheme~\cite{skb-nim2018} as in Fig.~\ref{skekb-config}. A beam emittance of about 20 mm.mrad was required with a charge of 4 nC per bunch. The positron capture section was reconstructed~\cite{fc-ipac2021} in order to increase the positron charge employing a flux concentrator pulsed solenoid, large aperture S-band structures (LASs), long DC solenoid magnets and a hundred of quadrupole focusing magnets. A damping ring was newly constructed to damp the positron emittance. For the electron beam, a new high-current and low-emittance RF gun was developed with an iridium-cerium photocathode and a quasi-travelling-wave side-coupled cavity (QTWSC)~\cite{rfgun-linac2014}. 

More than hundred pulsed focusing and corrector magnets were newly installed~\cite{pulmag-linac2018}.
Pulsed low-level RF controls were also developed. 
Based on those equipment upgrades~\cite{rejuve-ipac2018}, the simultaneous top-up injections for four storage rings of SuperKEKB HER, LER, PF ring and PF-AR were realized in 2019, 
which contributed to the considerable improvement of experimental efficiency~\cite{simul-pasj2020}. 

The continuous operation of the electron-positron injector has supported the history of advanced research at KEK by providing beams for various projects as shown in Fig.~\ref{projects}.

\begin{figure}[t]
\centering
\includegraphics[width=0.85\columnwidth]{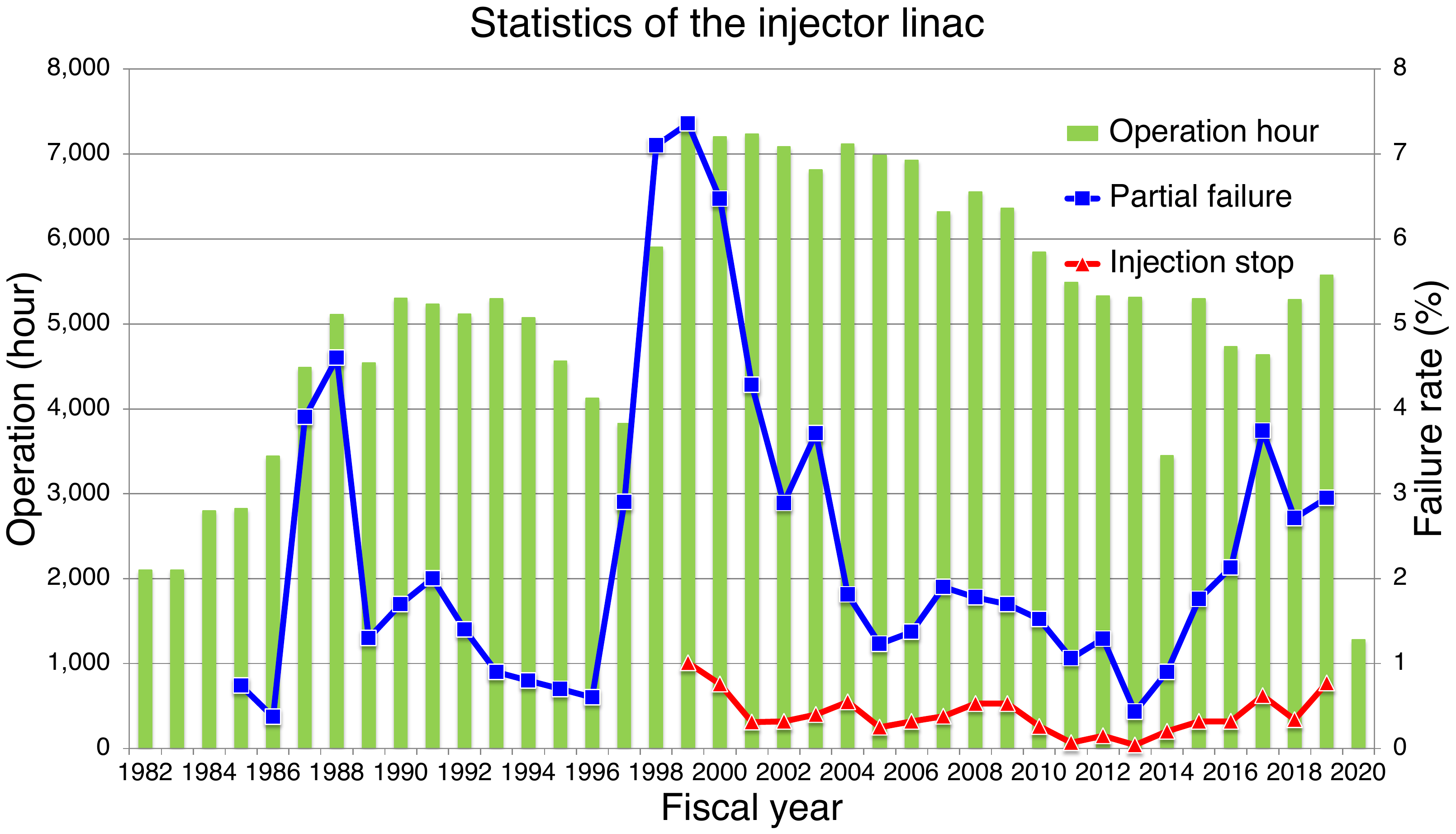}
\caption{
Yearly operation hours of the injector LINAC with failure rates.
}
\label{hours}
\end{figure}

At 8:50 a.m.\ on May 7, 2020, 200,000 hours of operation was accomplished as shown in Fig.~\ref{hours}, since the PF injection operation started in 1982. Authors are very proud to have reached the milestone of 200,000 hours of operation by inheriting the operation results accumulated by our predecessors. Although no special event was held due to the effects of COVID-19, a toast was made with morning coffee via video conference. 
In the statistics, the partial failure rate indicates that a certain device has failed but the beam injection operation was still possible due to redundancy, and the injection stop rate indicates that the failure was so serious that the beam injection was impossible. At the beginning of each program, the failure rate increases before understanding the equipment characteristics, but it tends to stabilize as countermeasures are taken.

\section{Conclusion}

The KEK electron-positron injector LINAC has been able to achieve 200,000 hours of operation while supporting experiments in multiple fields of photon science and particle physics. 



\section*{References}


\end{document}